\documentclass[onecolumn,superscriptaddress,prl,longbibliography]{revtex4-2}

\usepackage{graphicx}
\usepackage{bm}

\usepackage{xcolor}
\usepackage[colorlinks=true,citecolor=blue,linkcolor=magenta]{hyperref}
\usepackage{orcidlink}

\usepackage[latin9]{inputenc}
\usepackage{amsmath}
\usepackage{amssymb} 
\usepackage{graphicx}
\usepackage{verbatim}  
\usepackage{bm} 
\usepackage{color}  

\usepackage{graphicx,epsfig,amsfonts,amssymb}
\usepackage{bm}
\usepackage{times}
\usepackage{lipsum}
\usepackage{verbatim}

\newcommand{\la}{\langle}
\newcommand{\ra}{\rangle}

\newcommand{\bqe}{\begin{eqnarray}}
\newcommand{\eqe}{\end{eqnarray}}

\newcommand{\be}{\begin{equation}}
\newcommand{\ee}{\end{equation}}

\begin{document}
\title{Multivariable Painlev\'e-II equation: connection formulas for asymptotic solutions}
\author{Nikolai A. Sinitsyn\orcidlink{0000-0002-0746-0400}}\email{nsinitsyn@lanl.gov}
\affiliation{Theoretical Division, Los Alamos National Laboratory, Los Alamos, New Mexico 87545, USA} 
\begin{abstract}
For an integrable  generalization of the Painlev\'e-II equation (P-II) to a system of coupled equations with symmetry breaking terms,  an asymptotically exact WKB analysis is applied to obtain connection formulas for solutions at different infinities. The analysis relies on an exact solution of the quantum mechanical Demkov--Osherov model (DOM), revealing a possible deeper relation between classical integrable systems and solvable multistate Landau--Zener models. An application of the connection formulas to the problem of unstable vacuum decay during a second-order phase transition provides precise scaling of the number of excitations, including subdominant contributions.
\end{abstract}

\date{October 2025}
\maketitle

\section{Introduction}
Theoretical physics relies on a  set of special functions, which are usually solutions of differential equations whose properties can be understood in great detail. Thus, throughout the 20th century, a physicist's standard toolbox included knowledge of  solvable linear ordinary second-order differential equations. The commonly used property of such equations is the existence of {\it connection formulas} that relate the parameters of the asymptotic solutions in different limits.

Towards the end of the 20th century, the connection formulas for asymptotic solutions of certain  second order {\it nonlinear} differential equations were also derived.
For example, the homogeneous P-II
\be
u''(x)=xu(x)-2u(x)^3,
\label{P2-def}
\ee
is one of the most frequently encountered nonlinear second-order differential equations in theoretical physics. Connection formulas 
\cite{ItsKapaev1988ConnectionFormulasPII,Fokas2006} relate the behavior of $u(x)$ in the $x\to\pm \infty$ limits. They have been used to write integrals of Painlev\'e transcendents over $x$ in terms of classical special functions \cite{BaikBuckinghamDiFrancoIts2009,Kokocki2020,BothnerBuckingham2017}. Applications of such connection formulas have already been numerous, including in ultracold atoms \cite{Itin2009b,Sadhasivam2024}, viscous flow \cite{Lee2011ViscousShocks}, spectroscopy \cite{BanerjeeSinitsyn2023MesoscopicCriticalFluctuations}, quantum phase transitions \cite{SinitsynPokrovsky2026QuasiAdiabaticEffects}, quantum information \cite{ItsMezzadriMo2008}, plasma physics \cite{ZelenyiNeishtadtArtemyevVainchteinMalova2013, NeishtadtArtemyevTuraev2019}, liquid crystals \cite{ClercDavilaKowalczykSmyrnelisVidalHenriquez2017}, and stochastic processes \cite{BothnerLiechty2013}.

In a seemingly different direction, there has been a progress on solvable {\it higher-order linear} systems, including the multistate Landau--Zener models with linearly time-dependent parameters \cite{Sinitsyn2017b,Sinitsyn2016,Sinitsyn2018,Chernyak2019}  and solvable quantum quench models with decaying couplings \cite{Jeff-Coulomb,Li2018,Yan2022,Barik2025HigherSpinRG,PasnooriYuzbashyan2025,78xb-5lmw,cr3y-71lr}. They describe interacting quantum systems, including systems with combinatorial complexity, under explicitly time-dependent conditions.
This raises the question of whether there exist practically interesting higher-order {\it systems}  of nonlinear differential equations  that are integrable and possess simple connection formulas for the asymptotic behavior of their solutions as well.

The present Letter provides  exact analytical connection formulas for a multivariable generalization of P-II. This nonlinear system is  shown to be related to the Demkov--Osherov model (DOM) \cite{Demkov1967, SinitsynPokrovsky2026QuasiAdiabaticEffects}, which  is one of the earliest known solvable generalizations of the quantum mechanical Landau--Zener model to an arbitrary number of interacting states with linearly time-dependent parameters. As the class of solvable multistate Landau--Zener systems is  known to be large \cite{Chernyak2018,Chernyak2020}, its relevance to solvability of nonlinear systems suggests that the class of integrable multi-variable generalizations of the Painlev\'e equations with tractable asymptotic behavior may be as large. Thus, the synergy between the two research directions in mathematical physics may enable analytical treatment of more complex nonlinear problems.

\section{Definition of the model} 
Let a system of $n$ coupled nonlinear differential equations depend on $n$  parameters $\varepsilon_1<\varepsilon_2<\ldots <\varepsilon_n$:
\begin{eqnarray}
\nonumber u_1''(x)&=&xu_1(x)-2u_1(x)\sum_{k=1}^n u_k^2(x)- \varepsilon_1u_1(x),\\
\nonumber u_2''(x)&=&xu_2(x)-2u_2(x)\sum_{k=1}^n u_k^2(x)-\varepsilon_2 u_2(x),\\
\label{P2-n}
&\cdots& \\
\nonumber u_n''(x)&=&xu_n(x)-2u_n(x)\sum_{k=1}^n u_k^2(x)-\varepsilon_n u_n(x).
\end{eqnarray}
A shift, $x\rightarrow x+\varepsilon_1$,  sets $\varepsilon_1=0$, which will be assumed here.  

The system in Eq.~(\ref{P2-n}) is integrable. Without the linear $x$-dependence it is known as the Garnier system \cite{Garnier1919}, which is related to the multicomponent Korteweg--de Vries equation \cite{ChudnovskyChudnovsky1978}. 
The integrability of the system~(\ref{P2-n}) has been discussed in  Ref.~\cite{ClaeysDoeraene2018}, which mentioned that Eq.~(\ref{P2-n}) arises from the traveling-wave ansatz applied 
to the vector nonlinear Schr\"odinger equation with a linearly growing potential. Reference~\cite{ClaeysDoeraene2018}  also discussed
 applications of this system to stochastic processes and random matrix theory, and proved theorems about asymptotic solutions of Eq.~(\ref{P2-n}) that decayed exponentially with $x$. However, dynamical physical problems, such as the one in section~\ref{vac-dec},  require connection formulas for the most typical solutions -- of the Its--Kapaev type \cite{ItsKapaev1988ConnectionFormulasPII,Fokas2006} that can be parametrized by $3n$ free parameters, including $\varepsilon_k$ plus amplitudes and oscillation phases for all $u_k(x)$ as $x\rightarrow -\infty$. Such formulas did not appear in \cite{ClaeysDoeraene2018}. Moreover, the Riemann--Hilbert approach used in \cite{ClaeysDoeraene2018} may need considerable refinement to be adjusted to the Its--Kapaev boundary conditions \cite{Fokas2006}.
 
Here, the desired  asymptotic behavior of the solution as $x\rightarrow \pm \infty$ will be connected by explicit analytic formulas, written in the form convenient for applications to physical problems, as we will illustrate in later section~\ref{vac-dec}.  
The {\it first observation} that enables this result is that the system in Eq.~(\ref{P2-n}) is obtained as a consistency condition
\begin{equation}
\label{consHH}
\left(\frac{\partial H}{\partial x} - \frac{\partial H_1}{\partial t}\right) - i [H,H_1]=0,
\end{equation}
for two $(n+1)$-dimensional Hermitian for real $t$ and $x$ matrices, $H(t,x) $ and $H_1(t,x)$, given explicitly by  
\begin{eqnarray}
\label{HH-gen}
H &=& A - 4t B + 2C, \quad
H_1 \!=\! 
\left[\begin{array}{ccccc}
t & -u_1 & \cdots&-u_{n-1} & -u_n\\
-u_1 & -t & 0 &\cdots& 0\\
\vdots& \vdots& \ddots& \vdots& \vdots  \\
-u_{n-1} & 0 & \cdots &-t& 0\\
-u_n & 0 & \cdots & 0& -t
\end{array}\right],
\end{eqnarray}
where
\begin{eqnarray}
\nonumber &&A = \\
\nonumber &&
\small{ \left[\begin{array}{ccccc}
4t^2+x-2\sum \limits_{k=1}^nu_k^2(x) & 0 & \cdots & \cdots& 0\\
0 & -(4t^2+x-2\varepsilon_1)+2u_1^2 &  2u_1u_2 & \cdots  &2u_1u_n \\
\vdots &2u_2u_1 & -(4t^2+x-2\varepsilon_2)+2u_2^2& \ddots & 2u_2 u_n\\
\vdots & \vdots & \ddots & \ddots & \vdots\\
0 & 2u_nu_1(x) & 2u_nu_2 &\cdots &  -(4t^2+x-2\varepsilon_n)+2u_n^2
\end{array}\right],}\\
\nonumber \\
\nonumber \\
&&B =
\left[\begin{array}{cccc}
0 & u_1 & \cdots & u_n\\
u_1 & 0 & \cdots & 0\\
\vdots & \vdots & \ddots & \vdots\\
u_n & 0 & \cdots & 0
\end{array}\right],
\quad
C =
\left[\begin{array}{cccc}
0 & -i u_1' & \cdots & -i u_n'\\
i u_1' & 0 & \cdots & 0\\
\vdots & \vdots & \ddots & \vdots\\
i u_n' & 0 & \cdots & 0
\end{array}\right], \quad {\rm where } \quad   u_k' \equiv \frac{du_k(x)}{dx}.
\end{eqnarray}

Thus, if the functions $u_k(x)$, $k=1,\ldots n$, satisfy Eq.~(\ref{P2-n}), then $H$ and $H_1$ from Eq.~(\ref{HH-gen}) satisfy the consistency condition (\ref{consHH}). Similar properties have been previously used in studies of the  Painlev\'e equations \cite{Fokas2006,DomrinSuleimanov2025,AdlerSokolov2025} and many other nonlinear differential equations of mathematical physics \cite{Faddeev1987}.
Therefore, the fact that similar Lax pairs can be found for higher-order nonlinear equations  is not surprising. In fact, in Ref.~\cite{ClaeysDoeraene2018}, a different Lax pair was derived for the same system. The one presented here, however, is more convenient for the following WKB analysis, as it deals with Hermitian Hamiltonians having no singularities at finite real $x$ and $t$. Therefore, we can readily apply well-justified theoretical physics methods, without ever treating $t$ and $x$ as complex variables. 

Importantly, the consistency conditions usually do not guarantee that the WKB analysis that they enable leads to analytically tractable equations. The {\it second and main observation} that enables the following results is that such a WKB analysis for the system in Eq.~(\ref{P2-n}) can be performed completely analytically.

\section{The system of two variables: main results}

Derivation of the connection formulas for the entire system in Eq.~(\ref{P2-n}) is somewhat involved. The goal of this brief Letter is, instead, to announce their existence and demonstrate the consequences of the integrability for the simplest nontrivial case of only two variables, $n=2$ in Eq.~(\ref{P2-n}), while leaving many details to future publications.
For $n=2$, the system in Eq.~(\ref{P2-n}) reduces to
\begin{eqnarray}
\label{P2-1}
u_1''(x)&\!=\!&xu_1(x)\!-\!2u_1(x)\left[u_1^2(x)\!+\!u_2^2(x)\right],\\
\label{P2-2}
u_2''(x)&\!=\!&xu_2(x)\!-\!2u_2(x)\left[u_1^2(x)\!+\!u_2^2(x)\right]\!-\!\varepsilon u_2(x),
\end{eqnarray}
where  $\varepsilon>0$.

\subsection{Connection formulas}
Standard perturbative analysis \cite{Landau1980} fixes the asymptotic behavior of the solution for the system in Eqs.~(\ref{P2-1}) and (\ref{P2-2}) as $x\rightarrow -\infty$ up to four parameters, 
 $\alpha_{1,2}$ and $\varphi_{1,2}$, where $\alpha_{1,2}>0$. Namely,
\begin{eqnarray}
    \label{u1m}
   \nonumber     {\rm for} && x\rightarrow -\infty:\\
\nonumber    u_1&\!=\!&\frac{\alpha_1}{(-x)^{1/4}} \sin \left[ \frac{2}{3}(-x)^{3/2}+\frac{(3\alpha_1^2+2\alpha_2^2)}{4}\ln (-x) +\varphi_1 \right], \\
 u_2&\!=\!& \frac{\alpha_2}{(-x+\varepsilon)^{1/4}} \sin \left[\frac{2}{3}(-x+\varepsilon)^{3/2} + \frac{(3\alpha_2^2 +2\alpha_1^2)}{4}\ln (-x) +\varphi_2 \right].
\end{eqnarray}
The constant parameters $\alpha_{1,2}$ and $\varphi_{1,2}$ will be referred to as initial amplitudes and phases, respectively. 

As $x\rightarrow +\infty$, the trivial perturbation theory easily fixes only the leading $x$-dependent terms in the oscillation phases, while the sub-leading terms, starting with the terms behaving as $\propto \ln x$, are derived using the WKB approach described in Section~\ref{WKB-sec}. The solution is finally, i.e., as $x\rightarrow +\infty$, parametrized by two positive amplitudes, $\rho $ and $A$, two phases, $\phi_{1}$ and $\phi_2$, and one sign parameter, $\sigma=\pm 1$:
\begin{eqnarray}
\label{u1-largex}
 \nonumber  {\rm For}   &&  x\rightarrow +\infty:\\
u_1(x) &=& \sigma \sqrt{\frac{x}{2}} +\sigma \frac{\rho}{(2x)^{1/4}} \cos\left[ \frac{2\sqrt{2}}{3} x^{3/2} -\frac{3}{2}\rho^2\ln x +\phi_1\right],\\
\nonumber \\
 \label{u2-largex}
 u_2(x)&=&  A\cos\left[\sqrt{\varepsilon}x-\frac{A^2\sqrt{\varepsilon}}{2}\ln x+\phi_2\right].
\end{eqnarray}

Let me define the following combinations of the final and initial parameters:
\be
I_1\equiv \frac{\rho^2}{2}, \quad I_2\equiv \frac{A^2\sqrt{\varepsilon}}{2},
\label{actions-def}
\ee
\be
p_1=e^{-\pi \alpha_1^2}, \quad p_2\equiv e^{-\pi \alpha_2^2},
\label{inprob-def}
\ee
including two phases:
\begin{eqnarray}
\nonumber    \Phi_1&\equiv&\frac{\pi}{4}+{\rm arg} \Gamma\left( i\frac{\alpha_1^2}{2}\right) +\varphi_1-\frac{3\alpha_1^2}{2}\ln 2+\frac{\alpha_2^2}{2}\ln(\varepsilon/4),\\
\nonumber    \Phi_2 &\equiv& \frac{\pi}{4}+{\rm arg} \Gamma\left( i\frac{\alpha_2^2}{2}\right) +\varphi_2-\frac{3\alpha_2^2}{2}\ln 2+\frac{\alpha_1^2}{2}\ln(\varepsilon/4).
\end{eqnarray}
The {\it main result of this article} is the following connection formulas that relate the final parameters
$
\{\sigma, I_1,\, I_2, \phi_1, \phi_2 \}
$
to the set of initial parameters
$
\{p_1, p_2, \Phi_1, \Phi_2 \}
$ and the parameter of the equation, $\varepsilon$:

\begin{eqnarray}
\label{sigma-fin2}
 \sigma&=& {\rm sign}\left[ \sin \left(\Phi_1\right)\right],\\
 \label{rho-fin2}
  I_1&=&-\frac{1}{4\pi} \ln \left( 1-p_1^2p_2^2\large \left|1+\frac{1-p_1}{p_1}e^{2i\Phi_1} +\frac{1-p_2}{p_1p_2}e^{2i\Phi_2} \large\right|^2\right),\\
 \label{phi1-fin2}
  \phi_1&=&-\frac{3\pi}{4}- I_1 \cdot7\ln 2 +{\rm arg}\Gamma\left(2iI_1\right)-{\rm arg}  \left(1+\frac{1-p_1}{p_1}e^{2i\Phi_1} +\frac{1-p_2}{p_1p_2}e^{2i\Phi_2} \right),\\
  \label{I2-fin2}
  I_2&=&-\frac{1}{\pi} \ln \left(2\sqrt{p_2p_1(1-p_1)} \large \left| \sin \left(\Phi_1\right) \large \right| \right) -2I_1,\\
  \label{phi2-fin2}
  \phi_2&=&\frac{3\pi}{4}-\frac{2}{3}\varepsilon^{3/2}-I_2\ln(4\varepsilon^{1/2})+{\rm arg}\left[\Gamma(iI_2)\right]-{\rm arg}\left(e^{i\Phi_2}-e^{-i\Phi_2}\left(p_1+(1-p_1)e^{2i\Phi_1} \right)\right).
\end{eqnarray}

  \begin{figure}[t!]
    \centering
    \includegraphics[width=0.75\textwidth]{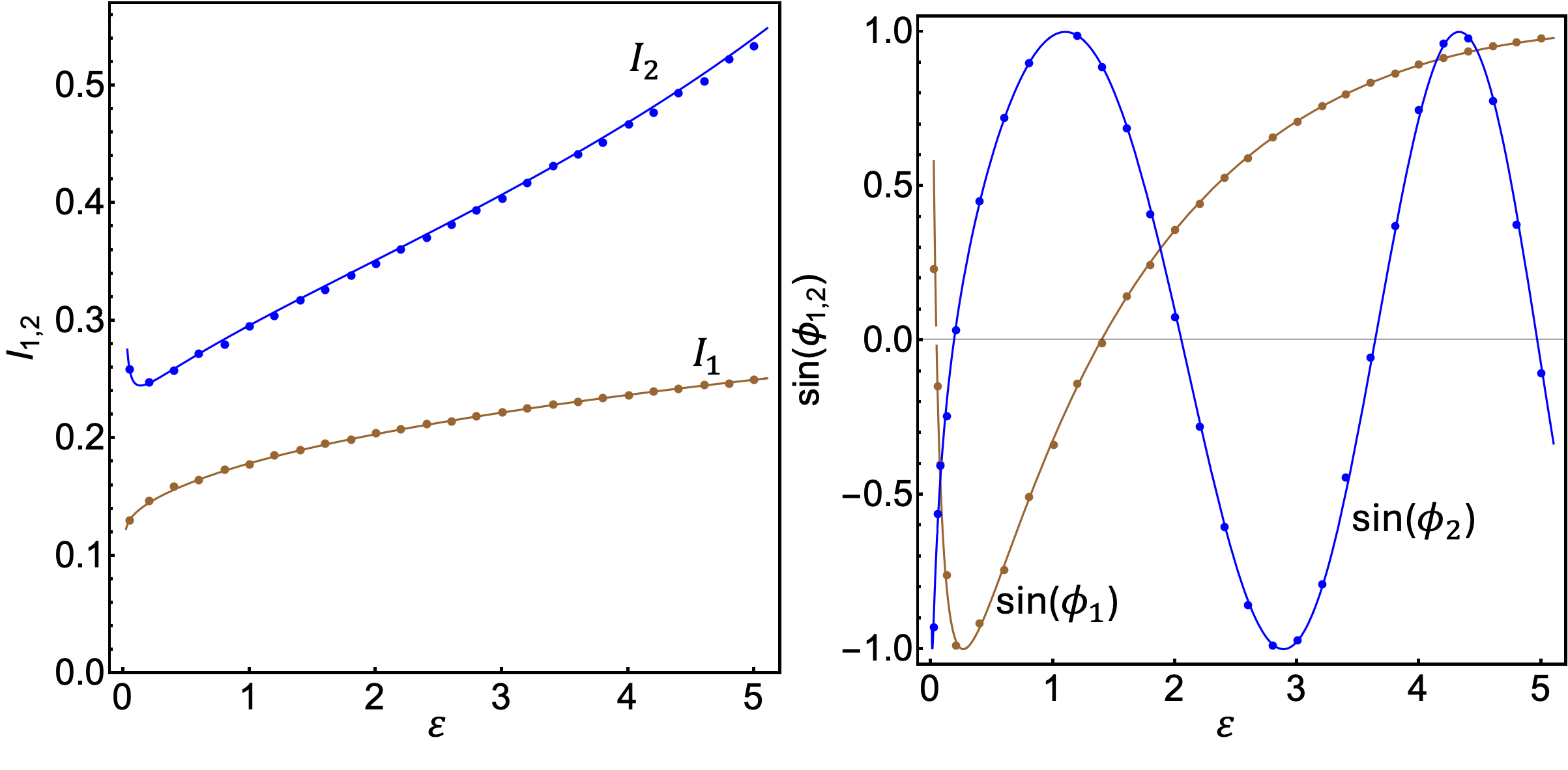}
    \caption{Numerical test of Eqs.~(\ref{rho-fin2})-(\ref{phi2-fin2}). Solid curves correspond to theoretical predictions for the dependence on the equation parameter $\varepsilon\in [0.05,5] $ (a) of the action variables $I_1$ (brown) and $I_2$ (blue); (b) of  the angle-related variables $\sin \phi_1$ (brown) and $\sin \phi_2$ (blue). Discrete points correspond to results of numerical simulations. Both for analytical predictions and numerical calculations the initial conditions were chosen to be $\alpha_1=0.9$, $\alpha_2=0.8$, $\varphi_1=\pi/2$, $\varphi_2=\pi/3$. Simulations were performed in the interval $x\in(-5000,5000)$ with a discretization step $dx=0.00001$. The algorithm is described in \cite{Tyagi2025}. Both theory and numerical simulations agree on that $\sigma=-1$ for all points here.
    }
    \label{teste-fig}
\end{figure} 

\begin{figure}[t!]
    \centering
    \includegraphics[width=0.9\textwidth]{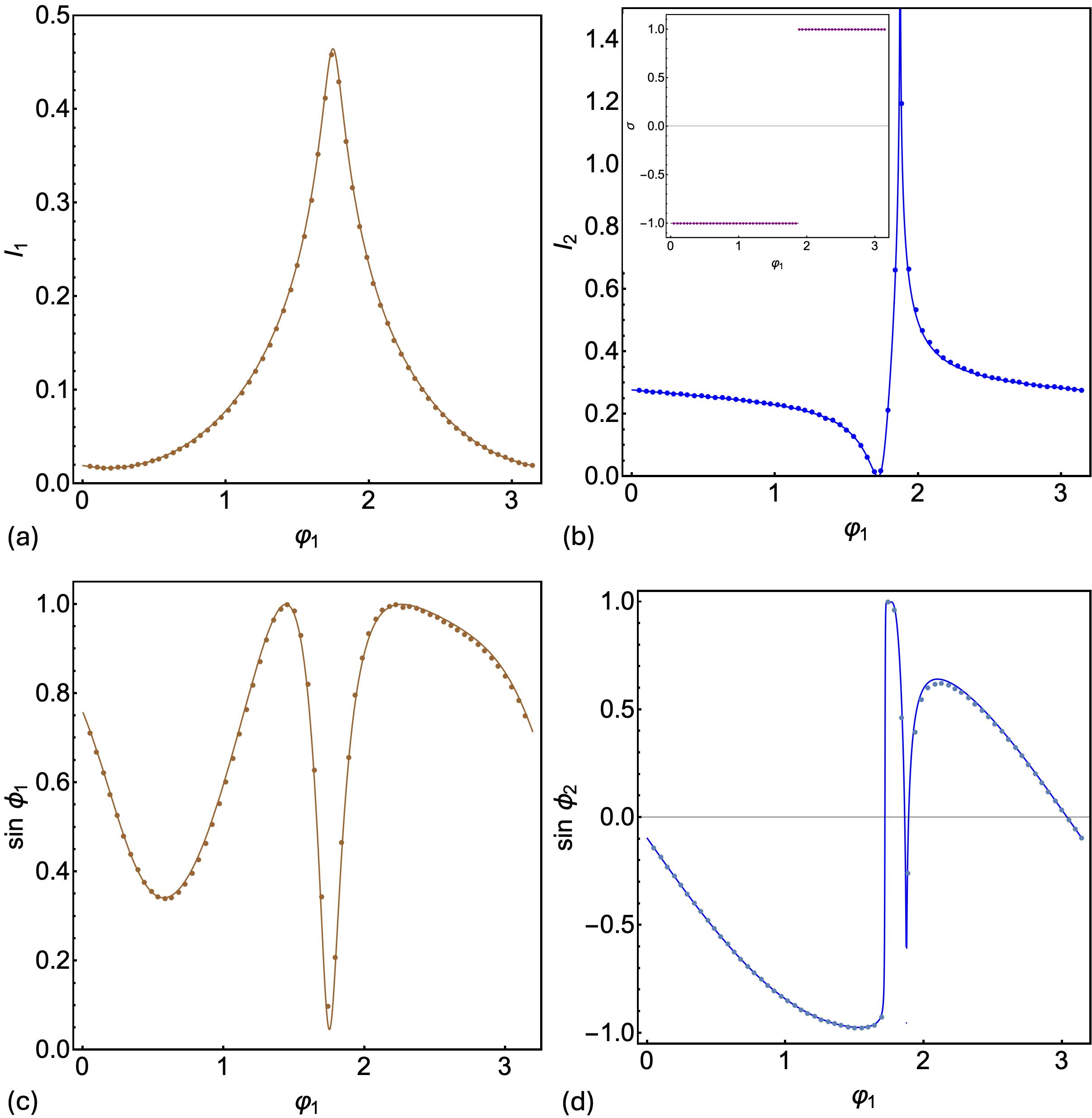}
    \caption{Numerical test of Eqs.~(\ref{sigma-fin2})-(\ref{phi2-fin2}). Solid curves correspond to theoretical predictions for the dependence on the angle $\varphi_1\in (0,\pi]$  (a,b) of the action variables $I_1$ (brown) and $I_2$ (blue), respectively; (c,d) of the angle-related variables $\sin \phi_1$ (brown) and $\sin \phi_2$ (blue), respectively. Discrete points correspond to results of numerical simulations. Both for analytical predictions and numerical calculations the initial conditions were chosen to be $\varepsilon=1$, $\alpha_1=0.8$, $\alpha_2=0.6$,  $\varphi_2=\pi/2$. Simulations were performed in the interval $x\in(-5000,5000)$ with a discretization step $dx=0.00001$. Inset in (b) shows the test of 
    Eq.~(\ref{sigma-fin2}) for the sign $\sigma$ dependence on $\varphi_1$ (purple).
    }
    \label{test-f1-fig}
\end{figure} 

 \subsection{Tests of connection formulas}
The first test for the validity of Eqs.~(\ref{sigma-fin2})-(\ref{phi2-fin2}) is the observation that at $\alpha_2=0$, Eq.~(\ref{P2-2}) has a trivial solution $u_2(x)=0$,
which corresponds to $I_2=0$ and $u_1(x)$ satisfying Eq.~(\ref{P2-def}). As expected, Eqs.~(\ref{rho-fin2}) and (\ref{phi1-fin2}) then reduce to the previously known connection formulas for Eq.~(\ref{P2-def}) listed in \cite{Fokas2006}.

More generally, the connection formulas were tested numerically for $x$-evolution with Eqs.~(\ref{P2-1}) and (\ref{P2-2}). Simulations were performed using the approach described in \cite{Tyagi2025}, which treats $x$ as a time variable. The evolution was considered over the interval $x\in (-5000,5000)$ with initial parameters taken to be $O(1)$. After this interval, a small piece of the trajectory was recorded and the final parameters were inferred by the best fit of this piece to Eqs.~(\ref{u1-largex}) and (\ref{u2-largex}).

The results are shown in Figs.~\ref{teste-fig} and \ref{test-f1-fig}. Since the angles $\phi_{1,2}$ are determined only up to an integer multiple of $2\pi$, to obtain smooth curves, $\sin \phi_1$ and $\sin \phi_2$ are plotted instead of the angles themselves.
Apparently, despite the highly nonlinear and sometimes singular dependence of the final set of parameters on the initial ones, the numerical results appear to be in perfect agreement with the analytical predictions. Many other similar tests (not shown) were performed for different values, e.g., of $\alpha_{1,2}$, with the same conclusion. 

Finally, let me comment on  the corrections to the connection formulas. For large but finite $x$, they appear starting with the terms $\propto 1/\sqrt{x}$, which decay relatively slowly. They can be calculated using the same WKB approach, but the complete study of such terms appears to be difficult due to the large number of diverse relevant contributions. Some of them are singular in the limit $\varepsilon \rightarrow 0$. This is why the simulations were restricted to $\varepsilon \ge 0.05$ -- otherwise, a considerably larger integration interval would be needed in order to achieve comparable agreement with the theoretical prediction.  

Numerically,  for the range of the tested parameters, with $\varepsilon \ge 0.05$, there were two specific corrections observed that dominated over the other terms of the same order in $x$. These two corrections can be easily included into the connection formulas, leading to noticeable accuracy improvements at large but finite $x$.
First, the regular part of $u_1(x)$ in Eq.~(\ref{u1-largex}) is renormalized by interactions as
\be
\sigma\sqrt{\frac{x}{2}} \rightarrow \sigma\sqrt{\frac{x-2u_2(x)^2}{2}},
\label{reg-ren}
\ee
where $u_{2}(x)$ is taken from Eq.~(\ref{u2-largex}). This correction vanishes as $\propto 1/\sqrt{x}$, but it is only a factor $\sim 1/x^{1/4}$ smaller than the leading oscillatory term in Eq.~(\ref{u1-largex}). Therefore, its effects are well visible during simulations as a slowly decaying modulation of oscillations in $u_1(x)$.

Second, there is a correction to $\phi_1$ in Eq.~(\ref{phi1-fin2}) that follows from the renormalization of the energy of the state $|0\ra$ in the WKB approach applied to the evolution along the path $P_{\infty}$ (see Section~\ref{WKB-sec} for details). It can be included in the connection formulas by subtracting an additional term from the right-hand side of Eq.~(\ref{phi1-fin2}):
\be
\phi_1 \rightarrow \phi_1-2\pi I_2\sqrt{\frac{\varepsilon}{2x}}.
\label{phi2-corr}
\ee
Although well suppressed, small effects of this correction were visible even for the longest simulations, so it was included in the comparison to numerical data for $\sin \phi_1$.  The relative importance of this correction is attributed to its growth with $\varepsilon$ and the existence of singularities that make $I_2$ anomalously large at specific initial conditions, as shown in Fig.~\ref{test-f1-fig}(b).   This singularity  follows from the existence of unstable solutions with exponential decay of $u_1(x)$, or both $u_1(x)$ and $u_2(x)$, for large $x$ at special values of the initial parameters. We postpone discussion of such unstable solutions, as they do not affect our analysis of a physical process in section~\ref{vac-dec}.

\section{Derivation of connection formulas}
\label{WKB-sec}
\subsection{Strategy}
Consistency conditions (\ref{consHH}) allow  definition of a state vector, $|\Psi(t,x) \ra$, as a solution of the two-time Schr\"odinger equation \cite{Faddeev1987}
\begin{eqnarray}
\label{SE1}
i\frac{\partial |\Psi\ra}{\partial t} &=& H(t,x) |\Psi \ra,\\
\label{SE2}
i\frac{\partial |\Psi\ra}{\partial x} &=& H_1(t,x) |\Psi \ra,
\label{multi-time}
\end{eqnarray}
where   $H$ and $H_1$  are  3$\times$3 matrices given by Eq.~(\ref{HH-gen}) for $n=2$.   Henceforth,  $H$ and $H_1$ will be called  {\it Hamiltonians} to stress that they are Hermitian operators for real values of $t$ and $x$, so that the
WKB techniques developed previously specifically for quantum mechanical problems can be utilized. 
The evolution operator $U_P$ along an arbitrary path $P$ in the two-time space $(t,x)$  can be written as a path-ordered exponent:
\be
U_P={\cal T}_{P} e^{-i\int_P \{H\,dt +H_1 \,dx \} },
\label{u-tx}
\ee
where ${\cal T}_{P} $ is the path-ordering operator, such that factors corresponding to earlier points along $P$ appear further to the right inside the product of evolution exponents $e^{-i\{H\,dt +H_1 \,dx \} }$. 

\begin{figure}[t!]
    \centering
    \includegraphics[width=0.45\textwidth]{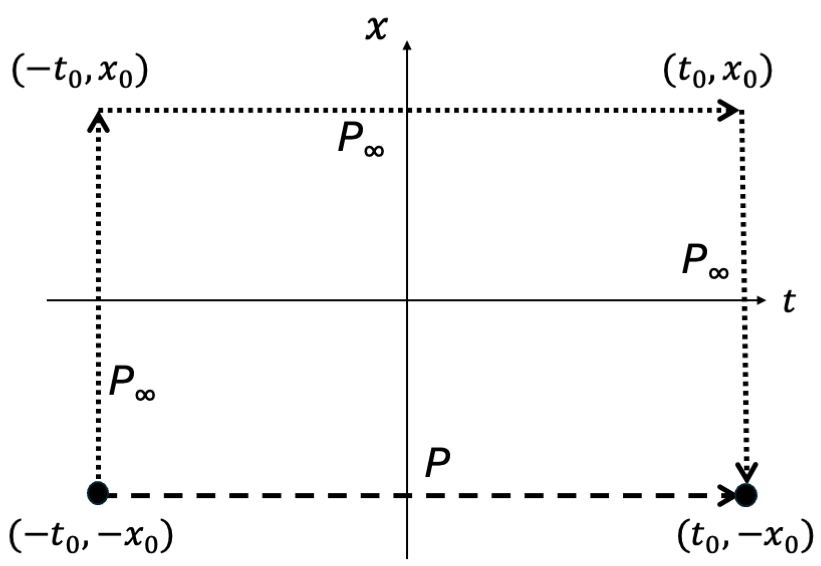}
    \caption{An integration path ${P}$ (dashed arrow) with $x_0\rightarrow -\infty$ and $t\in (-t_0,t_0)$, where $t_0\gg x_0$, is deformed into the path ${P}_{\infty}$, such that the horizontal segment of $ P_{\infty}$ lies at $x_0\rightarrow +\infty $ and
    $t\in (-t_0,t_0)$ (dotted arrows).  This deformation does not change the evolution operator in Eq.~(\ref{u-tx}), since the initial and final points of the path remain unchanged. The vertical legs of $P_{\infty}$ have $t=\pm t_0 \rightarrow \pm\infty$, which makes the evolution along them adiabatic. }
    \label{paths-fig}%
\end{figure}

 Equation~(\ref{u-tx}) can be written as 
\begin{equation}
U_P={\cal T}_{P} e^{\int_P {\bm A}\cdot d{\bm \tau}}, 
\label{gauge-exp}
\end{equation}
where ${\bm \tau} \equiv (t,x)$ is a two-time point and ${\bm A} \equiv (-iH,-iH_1)$ is a non-Abelian field. This field is flat (has zero curvature) \cite{Sinitsyn2018}, so the result of the evolution in Eq.~(\ref{gauge-exp}) depends only on the  endpoints, which are $(-t_0,-x_0)$ and  $(t_0,-x_0)$ in Fig.~\ref{paths-fig}, but does not depend on the choice of the path  connecting these points. 
 
\vspace{0.2cm}
Thus, the evolution over the path $P$ shown in Fig.~\ref{paths-fig} can be calculated in two ways. First,  one can consider $t$ changing in the interval $t\in (-t_0,t_0)$ at fixed negative $x=-x_0$. The evolution operator then takes the form 
\be
U_P={\cal T}_{t} e^{-i\int_{-t_0}^{t_0} H(t,-x_0)\,dt }.
\label{u-tx2}
\ee

Alternatively, $U_P$ can be obtained as the evolution operator along the path $P_{\infty}$ in Fig.~\ref{paths-fig}. The vertical legs of $P_{\infty}$ correspond to evolution at fixed $t$. For example, evolution along the left vertical leg is described by the  operator 
$$
U_{vl}={\cal T}_{x} e^{-i\int_{-x_0}^{x_0} H_1(-t_0,x)\,dx },
$$
where  the corner points of the paths are taken as $t_0,x_0 \rightarrow \infty$, with the limit $t_0 \rightarrow \infty$ taken first. 

Let $|k\ra$, where $k=0,1,2$, be the diabatic states, in which basis the Hamiltonians $H$ and $H_1$ are written in Eq.~(\ref{HH-gen}).  For $ t_0\rightarrow \infty$, these states coincide with the eigenstates of $H_1(-t_0,x)$. The evolution along
the left and right legs of $P_{\infty}$ is adiabatic,  resulting only in phase factors $e^{\pm 2it_0x_0}$. Thus, only the horizontal segment of $P_{\infty}$ contributes to the inter-state transitions. 

Both the horizontal part of $P_{\infty} $ and the path $P$ describe unitary quantum evolution over a pseudo-time $t\in (-t_0,t_0)$ at large positive and large negative $x$, respectively. Since the operator $H(t,x)$ depends on $u_{1,2}(x)$,    the relations between the asymptotic values of $u_{1,2}(x)$ as $x\rightarrow \pm \infty$ are obtained by equating $U_{P}$ to $U_{P_{\infty}}$.  Since the inter-state transition amplitudes are determined by the horizontal paths, at either large positive or large negative $x$, both evolution operators  can be evaluated using the WKB approach, which becomes exact in the limits $x\rightarrow \pm \infty$.

\subsection{WKB analysis in the limit $x\rightarrow -\infty$}
 In this case, there are two resonant regions for $t\in (-t_0,t_0)$ near 
 $
 t=\pm \sqrt{|x|}/2,
 $
where all nonadiabatic transitions occur. For large $|x|$, these resonances are well
 separated. By rescaling time 
\be
t=\tau \sqrt{|x|},
\label{t-resc1}
\ee
the Schr\"odinger equation (\ref{SE1}) is transformed to 
$
i\frac{d}{d\tau} |\Psi \ra =H(\tau) |\Psi \ra
$,
where

\begin{align}
\label{h3}
\noindent &H(\tau) =\\
\noindent \nonumber &\small{\left[\begin{array}{ccc}
|x|^{3/2}(4\tau^2-1)-2\sqrt{x}(u_1^2(x)+u_2^2(x))&-4|x|\tau u_1-2i\sqrt{|x|}u_1'& -4|x|\tau u_2-2i\sqrt{|x|}u_2'\\
-4|x|\tau u_1+2i\sqrt{|x|}u_1'& -|x|^{3/2}(4\tau^2-1) +2\sqrt{|x|}u_1(x)^2& 2\sqrt{|x|}u_1(x)u_2(x) \\
-4|x|\tau u_2+2i\sqrt{|x|}u_2' &2\sqrt{|x|}u_1(x)u_2(x)& -|x|^{3/2}(4\tau^2-1)+2\sqrt{|x|}(\varepsilon +u_2(x)^2)
\end{array}
\right].}
\end{align}

  \begin{figure}[t!]
    \centering
    \includegraphics[width=0.95\textwidth]{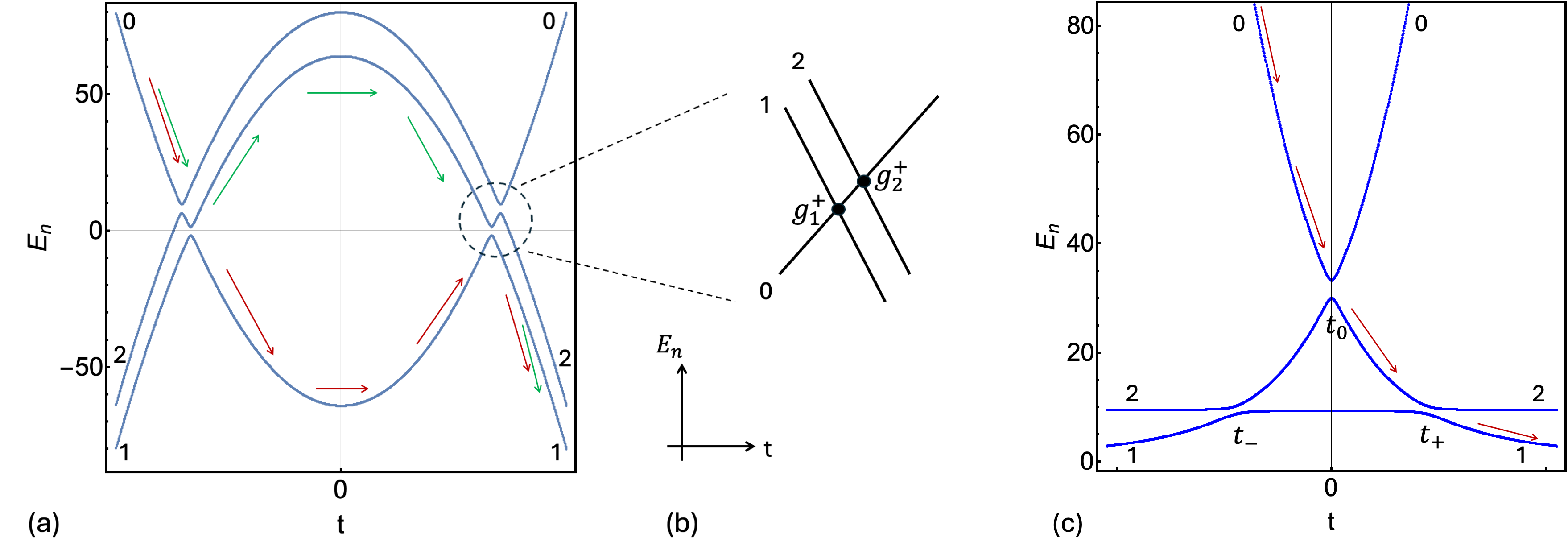}
    \caption{(a) Eigenvalues of the Hamiltonian~(\ref{h3}) as functions of $t=\sqrt{|x|}\tau$ at large negative $x$ (blue curves). Here, $x=-16$; $\alpha_1=\alpha_2=0.1$, $\varphi_1=\pi/3$, $\varphi_2=3\pi/4$. The values of $u_{1}(x)$ and $u_2(x)$ are approximated by asymptotic formulas in Eq.~(\ref{u1m}). The Hamiltonian~(\ref{h3}) is written in the basis of diabatic states.
The labels $0,\,1,\,2$ define the convention for the diabatic state indices. The diabatic states coincide asymptotically with the eigenstates as $t\rightarrow \pm \infty$. The red and green arrows show two semiclassical trajectories that originate on diabatic level $0$ as $t\rightarrow -\infty$ and terminate at level 1 as $t\rightarrow +\infty$.
The dashed circle encloses the behavior of the energy levels near the pseudo-time point $t=\sqrt{|x|}/2$. (b) Near $t=\sqrt{|x|}/2$, the dynamics is described by the exactly solvable DOM for three levels that cross linearly at fixed couplings $g_{1}^+$ and $g_2^+$. The diagram of diabatic levels shows only the time dependence of diagonal elements of the Hamiltonian, which in the DOM are straight lines in the time-energy plot. (c) Eigenvalues of the Hamiltonian $H(t,x)+x^{3/2}(4\tau^2+1)\hat{1}$,  as functions of $t=\sqrt{x}\tau$ at large positive $x$ (blue). Here, $x=10$, $A=0.2$, $\rho=0.12$, $\varepsilon=1.2$. A term proportional to $\hat{1}$ was added to the Hamiltonian to expose the spectrum better without changing its essential features.  The values of $u_{1}(x)$ and $u_2(x)$ are approximated by asymptotic formulas in Eqs.~(\ref{u1-largex}) and (\ref{u2-largex}), respectively. The red arrows show the unique path that connects diabatic levels $0$ and $1$. At large positive $x$, it passes through two avoided crossing points near times $t_0$ and $t_+$.
    }
    \label{DOspectr-fig}
\end{figure} 

The time-dependent spectrum of this Hamiltonian for large negative $x$ is shown in Fig.~\ref{DOspectr-fig}(a). Near $\tau=\pm 1/2$, additional simplifications follow.
Let
$$
g_{1,2}^-\equiv 2(|x|u_{1,2}-i\sqrt{|x|}u_{1,2}').
$$
After shifting the time variable to $\tau_m = \tau + 1/2$, the Hamiltonian in the vicinity of $\tau_m=0$ reduces to the three-state DOM Hamiltonian with one level crossing linearly two parallel levels:

\be
\label{pm1half}
\nonumber H_{-1/2}(\tau_m) = \left[\begin{array}{ccc}
-4|x|^{3/2}\tau_m&g_1^-& g_2^-\\
(g_1^-)^*& 4|x|^{3/2}\tau_m& 0 \\
(g_2^-)^* &0& 4|x|^{3/2}\tau_m+2\sqrt{|x|}\varepsilon
\end{array}
\right].
\ee


Similarly, consider the time interval near $\tau=1/2$ and keep only the relevant $x$-dependent terms. Introduce the couplings
$$
g_{1,2}^+=-2(|x|u_{1,2}+i\sqrt{|x|}u_{1,2}').
$$

After a shift of time as $\tau_p = \tau-1/2$, the Hamiltonian near $\tau_p=0$ is again the DOM:
\be
\label{hp1half}
\nonumber H_{1/2}(\tau_p) = \left[\begin{array}{ccc}
4|x|^{3/2}\tau_p&g_1^+& g_2^+\\
(g_1^+)^*& -4|x|^{3/2}\tau_p& 0 \\
(g_2^+)^* &0& -4|x|^{3/2}\tau_p+2\sqrt{|x|}\varepsilon
\end{array}
\right].
\ee
The diabatic level diagram for $H_{1/2}$ near the time-point $\tau_p=0$ is shown in Fig.~\ref{DOspectr-fig}(b). The two parallel levels may be close to one another, so that all three states interact simultaneously near the avoided crossings. Nevertheless, this model is solvable analytically.

The DOM exact solution for the scattering amplitudes coincides with the prediction of the independent crossing approximation, in which each encountered diabatic level crossing is treated as an independent two-state Landau--Zener transition between the crossing energy levels, followed by adiabatic evolution between successive level crossings \cite{SinitsynPokrovsky2026QuasiAdiabaticEffects}. Therefore, the entire calculation of the elements of the evolution matrix $U_{P}$ reduces to identifying semiclassical trajectories connecting the initial and final states and then finding the corresponding quantum amplitudes using the independent crossing approximation. An example of two semiclassical trajectories connecting the initial state $|0\ra$ to the final state $|1\ra$ is shown by the red and green arrows in Fig.~\ref{DOspectr-fig}(a).
The transition amplitudes between diabatic states are obtained  using the scattering matrix for the Landau--Zener model (listed, e.g., in Chapter 5 of Ref.~\cite{SinitsynPokrovsky2026QuasiAdiabaticEffects}) to describe transitions between pairs of states at any diabatic level intersection. Between such pairwise scatterings, the system undergoes adiabatic evolution with energies that can be estimated within second-order quantum perturbation theory in the small parameter $1/|x|$.

In the present case, the full evolution matrix $U_P$ is not required. It is sufficient to assume that initially the system is in state $|0\ra$, which has the highest energy as $t\rightarrow -\infty$, thus leading to the derivation of transition amplitudes $U_{P,00}$, $U_{P,10}$, and $U_{P,20}$.

This WKB approach has been used to derive some of the earliest connection formulas for Painlev\'e equations \cite{Novokshenov1984DE}. Here, due to multiple encountered avoided crossings of the Landau--Zener type, it becomes equivalent to the textbook independent crossing approximation, which is also used in theoretical physics as an approximation and for the derivation of transition amplitudes in multistate Landau--Zener models \cite{Chernyak2018} (author recommends Section~5.4 of Ref.~\cite{SinitsynPokrovsky2026QuasiAdiabaticEffects} for an elementary textbook introduction). The convenience of this approach is that the pseudo-time $t$ is never treated as a complex variable to obtain the evolution matrix elements. The accuracy of the independent crossing approximation has been well understood, so the method is believed to be well justified in theoretical physics. 
However, a mathematically more rigorous approach to deriving Painlev\'e asymptotic solutions is believed to be based on solving the Riemann--Hilbert problem \cite{Fokas2006}, which was not used in the present work, so  the connection formulas were validated by the numerical test.

\subsection{WKB analysis as $x\rightarrow +\infty$}
This limit is simpler because, as illustrated in Fig.~\ref{DOspectr-fig}(c), the level crossing pattern leads only to three elementary avoided crossings between pairs of levels. 
These crossings are well separated both in energy and in time $t$. For example, the times of diabatic level crossings are given by $t_0=0$ and $t_{\pm}\approx \pm x/(4\sqrt{\varepsilon})$. Therefore, each crossing is described by the scattering matrix for the two-state Landau--Zener model. Moreover, for the initial conditions with only level $0$ populated, the avoided crossing at $t_{-}$ is irrelevant, so only a single trajectory connects the initial state $|0\ra$ to any final diabatic state. Therefore, path interference does not occur. Figure~\ref{DOspectr-fig}(c) illustrates this property by showing 
the only semiclassical trajectory connecting levels $0$ and $1$. 

To calculate the scattering amplitudes using the asymptotically exact WKB method, the functions $u_{1,2}(x)$ as $x\rightarrow +\infty$ were assumed to take the form given in Eqs.~(\ref{u1-largex}) and (\ref{u2-largex}) with unknown parameters, including at the logarithmic phase contributions. Near each avoided crossing, the couplings were expressed in terms of these unknown parameters, and the scattering amplitudes were estimated using the known scattering matrix for the Landau--Zener model.

After obtaining the amplitudes of $U_{P,k0}$, where $k=0,1,2$, separately for  $P$ and $P_{\infty}$, the  WKB results were compared order by order in powers of $x$ and $t_0$, down to $O(1)$ contributions to the phases and the absolute values of the transition amplitudes. Terms that depended on $t_0$ and on high powers of $x$ on both sides then canceled, leaving the connection formulas written in Eqs.~(\ref{sigma-fin2})-(\ref{phi2-fin2}) and fixing the logarithmic $x$-dependence in Eqs.~(\ref{u1-largex}) and (\ref{u2-largex}).

\section{Excitations after   unstable vacuum decay }
\label{vac-dec}

The system in Eqs.~(\ref{P2-1}) and (\ref{P2-2}) was previously introduced in \cite{Tyagi2025,Suzuki2025} as describing a passage through a phase transition in scalar field theory.
It can be interpreted as Hamiltonian dynamics, in which $x$ is associated with physical time, $t$ (not to be confused with the pseudo-time introduced in the previous section), and $u_{1}(t)$ and $u_2(t)$ are  coordinates of a particle with two degrees of freedom. Let ${\bm X}=(u_1,u_2)$, and the corresponding time-dependent Hamiltonian is
\begin{equation}
{\cal H}(t)=\frac{{\bm P}^2}{2} -t \frac{{\bm X}^2}{2} +\frac{{\bm X}^4}{2} +  \frac{\varepsilon u_2^2}{2} ,
\label{hamu}
\end{equation}
where ${\bm P}=(P_1,P_2)$ is the vector of momenta canonically conjugated to $u_1$ and $u_2$, and where ${\bm X}^2\equiv u_1^2+u_2^2$. The system in Eqs.~(\ref{P2-1}) and (\ref{P2-2}) follows from  Newton's equations of motion with the Hamiltonian (\ref{hamu}).

In the limits $t\rightarrow \pm \infty$, the potential energy for this classical motion is that of a harmonic oscillator, so the amplitudes $\alpha_{1,2}$ are related to the asymptotically conserved, as $t\rightarrow -\infty$, adiabatic invariants \cite{ZelenyiNeishtadtArtemyevVainchteinMalova2013,SinitsynPokrovsky2026QuasiAdiabaticEffects}:
$$
{\cal I}_{1,2}\equiv \frac{\alpha_{1,2}^2}{2}.
$$
Similarly, $I_1$ and $I_2$ from Eq.~(\ref{actions-def}) are identified as asymptotically conserved adiabatic invariants as $t\rightarrow +\infty$. 

According to \cite{Tyagi2025,Suzuki2025}, this classical dynamics appears as a saddle-point solution for a quantum field theory describing a time-dependent passage through a second-order quantum phase transition. The difference between ${\cal I}_{1,2}$ and $I_{1,2}$ captures inevitable nonadiabatic effects as the classical phase space trajectory passes through a separatrix. Physically, this event corresponds to crossing the critical point of a phase transition in the corresponding field theory \cite{Suzuki2025}. Semiclassically, the actions are  related to the numbers of produced quasiparticles: $I_{1}/\hbar$ for the number of Higgs bosons and $I_{2}/\hbar$ for the number of Goldstone bosons. The latter acquire a small mass due to the symmetry breaking term $\propto \varepsilon$ in Eq.~(\ref{hamu}). The analysis in \cite{Tyagi2025} did not provide explicit connection formulas but made several observations that can now be verified.

Thus, a useful feature of Eqs.~(\ref{rho-fin2}) and (\ref{I2-fin2}) is that $I_{1}$ and $I_2$ depend on  $\varepsilon$ only through the angles $\Phi_{1}$ and $\Phi_{2}$,  which themselves depend linearly on the initial phases $\varphi_{1}$ and $\varphi_{2}$. Therefore, $\la I_{1}\ra$ and $\la I_{2}\ra$ are independent of $\varepsilon$,  where the brackets denote averaging over a uniform distribution of $\varphi_{1}$ and $\varphi_2$. This invariance of the averaged actions was conjectured in \cite{Tyagi2025} based on less rigorous analytical arguments and numerical simulations, and is now confirmed.

Next, the solution in Eqs.~(\ref{sigma-fin2})-(\ref{phi2-fin2})  exhibits asymmetry amplification, which was the main finding in \cite{Tyagi2025}. No matter how small the symmetry breaking parameter $\varepsilon$ is, the dynamics for $u_1(t)$ and $u_2(t)$  strongly differ from one another as $t\rightarrow +\infty$. While $u_1(t)$ has a monotonically growing component, the asymptotic solution for $u_2(t)$ oscillates around zero.
Note  that if $\varepsilon$ changes sign, the asymptotic solutions in Eqs.~(\ref{u1-largex}) and (\ref{u2-largex}) are obtained by switching indices $1$ and $2$. The solution at $\varepsilon=0$ reduces to the standard P-II solution, which is unstable in the present context.

The study  in \cite{Tyagi2025} also addressed the following question: Given that initially the system is nearly at the quantum vacuum state, which  corresponds, in  dimensionless variables here, to
\be
{\cal I}_1={\cal I}_2\equiv {\cal I} \ll 1,
\label{calJin}
\ee
what are the values of the adiabatic invariants as $t\rightarrow +\infty$? Averaging over the initial phases $\varphi_{1}$ and $\varphi_2$ was required.

Using that $\int_0^{2\pi} \ln(2|\sin(\Phi)|)\, d\Phi=0$, averaging Eq.~(\ref{I2-fin2}) over
$\varphi_{1,2}$ at conditions in Eq.~(\ref{calJin}) gives that 
\be
\la I_2\ra +2\la  I_1 \ra \approx \frac{1}{2\pi} \ln\left( \frac{1}{2\pi\cal I} \right).
\label{av-res1}
\ee
Similarly,  Eq.~(\ref{rho-fin2}) with Eq.~(\ref{calJin}) lead to
\be
\la I_1 \ra\approx \frac{1}{4\pi}\left[ \ln \left(\frac{1}{2\pi {\cal I}}\right) -c_1 \right],
\label{i1-num}
\ee
where
\begin{equation}
\label{c1-num}
 c_1=\frac{1}{\pi^2}\iint_0^{\pi} \ln \left[4-2\cos(2\Phi_1) -2\cos(2\Phi_2) \right]\,d\Phi_1d\Phi_2\approx 1.166.
\end{equation}
Substituting Eqs.~(\ref{i1-num}) and (\ref{c1-num}) into Eq.~(\ref{av-res1}) leads also to 
\be
\la I_2 \ra \approx \frac{c_2}{2\pi}, \quad c_2=c_1\approx 1.166.
\label{i2-num}
\ee

Expressions for $\la I_{1}\ra$ and $\la I_2 \ra$ in Eqs.~(\ref{i1-num}) and (\ref{i2-num}) were conjectured in \cite{Tyagi2025} for ${\cal I} \ll 1$ without an analytical derivation of $c_{1,2}$, while numerical simulations in \cite{Tyagi2025} estimated that $c_1\approx 1.14$ and $c_2\approx 1.19$, with an uncertainty in the last significant digit. This agrees well with the analytically derived values of these coefficients in Eq.~(\ref{i2-num}). Curiously, the value of the double phase integral in Eq.~(\ref{c1-num}), $ \approx 1.166$, is the famous square-lattice spanning tree constant \cite{PhysRevE.95.062138}.

\section{Conclusion}
When a system of nonlinear equations emerges from consistency conditions (\ref{consHH}) for matrix operators, the nonlinear problem is effectively described by a linear one. The difficulty for higher-order systems of nonlinear equations often lies in the lack of a fully analytical solution of the associated linear model, even when WKB methods are applicable. Nevertheless, the class of known fully solvable multistate linear systems has been growing recently. Any of the newly solved linear systems may arise in the WKB analysis enabled by the consistency conditions for a certain nonlinear system. The present article provides an example of such a situation, leading to analytical connection formulas for asymptotic solutions of two coupled P-II equations with a minimally broken symmetry.

\begin{acknowledgements}
Author thanks Prof.~Alexander~Its for useful discussions. This work was carried out under the
auspices of the U.S. DoE through the Los Alamos National Laboratory, operated by Triad
National Security, LLC (Contract No. 892333218NCA000001). 
\end{acknowledgements}
\bibliography{ref2}
\end{document}